\documentclass[aps,pre,amsmath,amssymb,reprint]{revtex4-1}
\usepackage{hyperref}
\usepackage{graphicx}
\usepackage{dcolumn} 
\usepackage{color}
\usepackage{grffile}
\usepackage{bm} 
\usepackage{blindtext}

\definecolor{apsblue}{rgb}{0.18,0.19,0.57}
\definecolor{darkblue}{rgb}{0.2,0.1,0.5}
\definecolor{darkgreen}{rgb}{0.1,0.6,.1}
\definecolor{darkred}{rgb}{0.7,0.0,.1}

\hypersetup{
 colorlinks,
 citebordercolor={0.3 0.8 .1},
 linkbordercolor={0.9 0.2 .1},
 urlbordercolor={0.6 0.6 0.8},
 citecolor=apsblue,
 urlcolor=apsblue,
 linkcolor=apsblue,
 pdfpagemode={UseNone},
 pdfauthor={},
 pdftitle={},
}

\graphicspath{{figures/}:{./}:{/home/coslo/projects/polydisperse_swap/images/}}

\begin{document}

\title{Local order and crystallization of dense polydisperse hard spheres}
\author{Daniele Coslovich}
\author{Misaki Ozawa}
\author{Ludovic Berthier}
\affiliation{Laboratoire Charles Coulomb, 
Universit\'e de Montpellier, CNRS, Montpellier 34095, France}

\begin{abstract}
Computer simulations give precious insight into the microscopic behavior of supercooled liquids and glasses, but their typical time scales are orders of magnitude shorter than the experimentally relevant ones. We recently closed this gap for a class of models of size polydisperse fluids, which we successfully equilibrate beyond laboratory time scales by means of the swap Monte Carlo algorithm. In this contribution, we study the interplay between compositional and geometric local orders in a model of polydisperse hard spheres equilibrated with this algorithm. Local compositional order has a weak state dependence, while local geometric order associated to icosahedral arrangements grows more markedly but only at very high density. We quantify the correlation lengths and the degree of sphericity associated to icosahedral structures and compare these results to those for the Wahnstr\"om Lennard-Jones mixture. Finally, we analyze the structure of very dense samples that partially crystallized following a pattern incompatible with conventional fractionation scenarios. The crystal structure has the symmetry of aluminum diboride and involves a subset of small and large particles with size ratio approximately equal to 0.5.
\end{abstract}

\maketitle

\section{Introduction}
\label{sec:intro}

Computer simulations play an important role in the study of amorphous materials, since they provide particle-scale resolution of their structural and dynamical properties. There is however a huge gap between the timescales  accessible in conventional simulations and in experiments on molecular and polymeric liquids. Despite the continuous increase of computing power, simulations timescales are still about $8$ orders of magnitudes shorter than experimental ones. Numerical analysis is therefore limited to studies of moderately supercooled liquids or poorly annealed glasses~\cite{berthier2011theoretical,binder2011glassy}.

Recently, we have developed a very efficient simulation setup by applying the swap Monte Carlo algorithm~\cite{gazzillo1989equation,frenkel2001understanding,grigera2001fast} to realistic models of polydisperse particles~\cite{ninarello2017,berthier2016equilibrium,berthier2017configurational}. 
Some of these models can be equilibrated even beyond the timescales accessible in the laboratory. Thanks to this simulation approach, it becomes possible to scrutinize under experimentally relevant conditions several outstanding issues concerning glass 
formation, such as the entropy crisis~\cite{berthier2017configurational}, the kinetic stability of ultrastable glasses~\cite{fullerton2017density,berthier2017origin}, jamming~\cite{ozawa2017exploring}, and the Gardner transition~\cite{berthier2015growing,jin2016exploring,scalliet2017absence}. These aspects are central in the current debate on the thermodynamic and dynamical properties of amorphous materials.

Local structure is another feature that may provide important insight into the thermodynamic and dynamic behavior of glass-formers~\cite{tanaka2012bond,Royall_Williams_2015}. Multi-component mixtures are characterized by local ``compositional'' order, which emerges due to preferential interactions between different chemical species~\cite{bhatia1970structural}. Systems with continuous polydispersity might have even more complex forms of compositional ordering~\cite{wilding2004phase,sollich2010crystalline,wilding2010phase}. A large body of experimental and simulation studies further demonstrated that simple glass-formers, such as colloids~\cite{leocmach2012roles}, metallic glasses~\cite{hirata2013geometric} as well as simple simulation models~\cite{coslovich2007understanding,soklaski2016locally,royall2017locally}, display a tendency to form locally favored structures as temperature decreases or density increases. 
The symmetry of these local structures is often  incompatible with the one of the underlying crystalline ground state~\cite{lechner2008accurate}, either because of compositional~\cite{Crowther_Turci_Royall_2015} or geometric frustration~\cite{tarjus2005frustration}.
In other cases, however, the preferred local order is the crystalline one~\cite{tanaka2012bond}, but the latter competes with an alternate local structure. The influence of this ``geometric'' local order on the dynamics of supercooled liquids~\cite{coslovich2007understanding,hocky_correlation_2014} and on their rheological properties~\cite{pinney2016structure,feng2015atomic,ding2014soft} has been the focus of several numerical studies. However, these studies were limited to the moderately supercooled regime and the ultimate role of local structure in the overall picture of glass formation is still under debate~\cite{tarjus2011overview,charbonneau2013decorrelation,Royall_Williams_2015}.

One outstanding issue of the local structure description is that the spatial correlations associated to  the geometric order are fairly small. The correlation lengths associated to locally favored structures remain small in the range of temperature and density accessible to conventional simulations~\cite{malins_identification_2013,malins2013lifetimes}. This behavior contrasts with the apparent increase of dynamic correlations, as measured from time-dependent multi-point functions~\cite{berthier2011theoretical}. These discrepancies might be attributed to model dependence~\cite{hocky_correlation_2014} or to the existence of different dynamic regimes not covered by standard simulations~\cite{royall2015strong}, but also raise some doubts about the physical relevance of local geometric order in the process of glass formation.

In this paper, we address these issues by analyzing a polydisperse fluid equilibrated very deeply with the swap Monte Carlo algorithm. We carefully analyze the role of compositional fluctuations and identify the preferred geometric motif of the system. We find that local compositional order increases smoothly  with increasing density. On the other hand, the geometric order associated to the preferred icosahedral order starts growing markedly only at sufficiently large volume fractions. We extract the correlation lengths associated to icosahedral structures and compare these results with a representative binary Lennard-Jones mixture. We further elucidate the interplay between compositional and geometric order in the polydisperse system. Finally, we analyze a partially crystallized sample that we obtained during long swap Monte Carlo simulation and rule out a conventional fractionation scenario for the model at hand. Overall, our results show that size polydisperse systems represent good glass-formers that are difficult to crystallize over a broad dynamical range, and are characterized by only weak static compositional fluctuations. By comparison with earlier models of glass-formers, they appear to contain much less local order at equivalent degree of supercooling. 

This paper is organized as follows. In Section~\ref{sec:methods} we present the numerical methods we use. In Section~\ref{sec:results} we present the results, which we organise into compositional order (Section~\ref{sec:compo}), geometric order (Section~\ref{sec:geo}), followed by an analysis of the crystal structure occasionally found in long simulations (Section~\ref{sec:xtal}). 
We conclude the paper in Section~\ref{sec:conclusions}.  

\section{Methods}
\label{sec:methods}

We study systems composed of $N$ polydisperse additive hard spheres of diameter $\sigma$ in three dimensions. The diameter distribution is the same as in Ref~\cite{berthier2016equilibrium}, $P(\sigma)=A\sigma^{-3}$, $\sigma_\textrm{min}\le \sigma\le \sigma_\textrm{max}$ with $\sigma_{\rm min}/\sigma_{\rm max}=0.4492$, where $A$ is a normalization constant. We use the average diameter $\overline{\sigma}=\frac{1}{N} \sum_{i=1}^N \sigma_i$ as the unit of length. In the following we mostly focus on samples of $N=8000$ particles, but we also carried out simulations for $N=64000$ to check for finite size effects. The simulations were performed using the swap Monte Carlo algorithm~\cite{gazzillo1989equation,sindzingre1989calculation,grigera2001fast} using the same setup as in~\cite{berthier2016equilibrium,berthier2017configurational}. This simulation approach is extremely efficient and allows one to equilibrate the fluid at least as deeply as conventional laboratory experiments on molecular liquids~\cite{ninarello2017}. We note that this is enabled by the combined optimization of both the Monte Carlo algorithm and the model parameters, which must be chosen such that the system is robust enough against crystallization or phase separation~\cite{ninarello2017}. 
Typical reference volume fractions of the system are onset of two-step relaxation ($\phi_{\rm onset} \simeq 0.56$) and mode coupling crossover ($\phi_{\rm mct} \simeq 0.6$). Current conventional simulations can equilibrate the fluid up to around $\phi_{\rm mct}$~\cite{brambilla2009probing,zaccarelli2015polydispersity}.
The initial configurations were prepared by fast Monte Carlo compressions of a low density fluid~\cite{Berthier_Witten_2009}, which was subsequently equilibrated at the target packing fraction.
We have checked that the configurations analyzed in the following correspond to an equilibrium, disordered fluid, by carefully
monitoring possible signs of crystallization or phase separation~\cite{wilding2004phase,wilding2010phase} using the same structural tools described
in~\cite{berthier2016equilibrium,berthier2017configurational}. 
One smaller sample of $N=1000$ particles showed clear signs of partial crystallizationous during long simulations at a high volume fraction ($\phi=0.648$). The structure of this sample will be analyzed separately in Section~\ref{sec:xtal}.

To probe the spatial structure of the system, we use generalized structure factors
\begin{eqnarray}
 S_w(k) &=& \frac{1}{N}  \langle \delta \rho_w({\bf k}) \delta \rho_w(-{\bf k}) \rangle, \\
\delta \rho_w({\bf k}) &=& \rho_w({\bf k}) - \langle \rho_w({\bf k}) \rangle,
\end{eqnarray}
where $\langle \cdots \rangle$ is the statistical average and $\rho_w({\bf k})$ is the Fourier transform of a weighted microscopic density
\begin{equation}
  \rho_w({\bf k}) = \sum_j w_j \exp{(-i {\bf k} \cdot {\bf r}_j)}.
  \label{eqn:rho_w}
\end{equation}
Here, the field $w_j$ is a generic particle property and enters as a weight in the calculation of the structure factor. In the following, we will consider various fields $w_j$.

The structure of simple mixtures and polydisperse particle systems is often characterized by some preferred local arrangements, also known as locally favored structures~\cite{Royall_Williams_2015}. To identify this kind of geometric local order we perform a radical Voronoi tessellation using the voro++ package~\cite{voro++}. In this construction, the total volume is partitioned into cells surrounding each particle in the system. Cells are then classified according to their signature $(n_3, n_4, n_5, \dots)$, where $n_q$ is the number of faces of the cell with $q$ vertices. Icosahedral local structures correspond to cells with the $(0,0,12)$ signature. In the following, we will further distinguish between particles that are at the center of an icosahedral structure and icosahedral structures as a whole~\cite{coslovich2007understanding}. A cluster of neighboring icosahedral centers will be called ``backbone'', while a cluster of neighboring icosahedral structures will be called ``domain'', see Section~\ref{sec:geo}.

\section{Results}
\label{sec:results}

\subsection{Compositional order}
\label{sec:compo}

\begin{figure*}[!ht]
  \centering
  \includegraphics[width=0.64\linewidth]{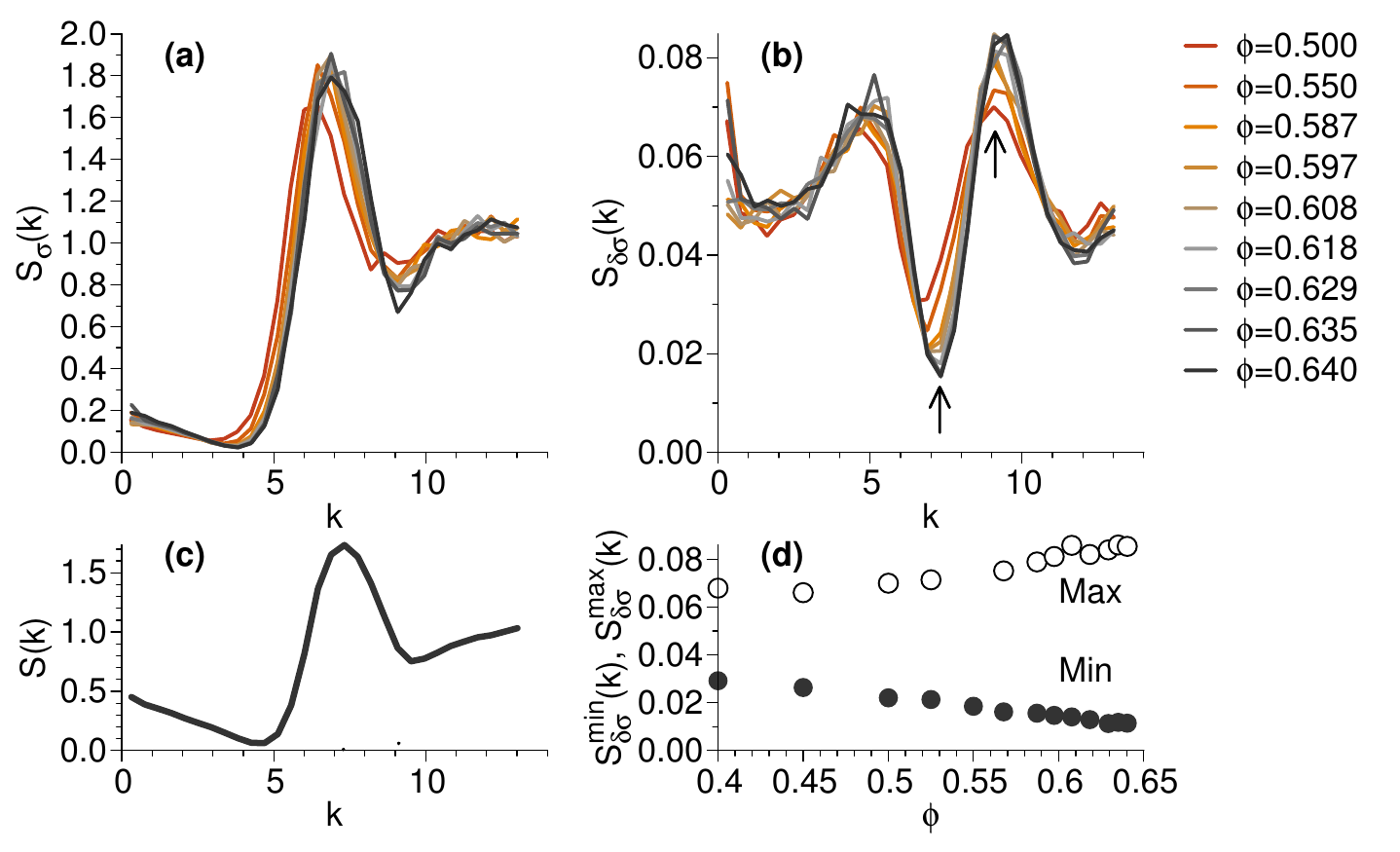}
  \caption{\label{fig:skvolume}
    (a) Structure factor of the diameter field $S_\sigma(k)$. (b) Structure factor of the diameter fluctuations field $S_{\delta \sigma}(k)$.
    (c) Total structure factor of the density field $S(k)$ at $\phi=0.64$. (d) Absolute minimum and absolute maximum values (specified by the arrows in (b)) of $S_{\delta\sigma}(k)$ as a function of $\phi$.}
\end{figure*}

In simple mixtures of particles, compositional order reflects the tendency of particles to coordinate according to their chemical species. A typical example is the presence of chemical short range ordering in metallic alloys, which reflects the tendency of particles to be surrounded by neighbors of a different chemical species~\cite{na2014compositional}. This effect can be quantified by computing partial structure factors and their linear combinations~\cite{bhatia1970structural}. This amounts to assigning a weight $w_i$ equal to $1$ or $0$ depending on whether particle $i$ belongs to a given species or not.

In a polydisperse system, the relevant microscopic weight associated to compositional order varies continuously. Since our system is size-dispersed, we are actually interested in the spatial structure associated to the diameter field. The simplest two-point correlation function that captures the local fluctuations of $\sigma_i$ is the diameter structure factor $S_\sigma(k)$, defined by setting $w_i = \sigma_i$ in Eq.~\eqref{eqn:rho_w}.
In Figure~\ref{fig:skvolume}, we show $S_\sigma(k)$  for several volume fractions ranging from the moderately dilute regime ($\phi=0.5$) to highly packed configurations ($\phi=0.64$). The correlation function varies smoothly and weakly as the system gets denser. Overall the shape of $S_\sigma(k)$ strongly resembles the one of the total structure factor $S(k)$~\cite{ozawa2017exploring}, shown in Figure~\ref{fig:skvolume}(c) at the largest volume fraction.

Since the diameter field by itself is weakly coupled to the local structure, we consider instead the structure factor $S_{\delta\sigma}(k)$ associated to the fluctuating part of the diameter field. To this end we use $w_i = \delta \sigma_i = \sigma_i - \overline{\sigma}$, where $\overline{\sigma}$ is the average particle diameter. We note that $S_{\delta\sigma}(k)$ is related to $S_\sigma(k)$ in a non-trivial way because of the presence of cross terms
\begin{equation}
S_{\delta\sigma}(k) = S_{\sigma} (k) + \overline{\sigma}^2 S(k) - \frac{2\overline{\sigma}}{N} \textrm{Re}[{\langle \delta \rho_{\sigma}({\bf k})\delta\rho(-{\bf k})\rangle}],
\end{equation}
where $\textrm{Re}[(\cdots)]$ is the real part of $(\cdots)$ and $\rho({\bf k})$ is the Fourier transform of the microscopic density with $w_j=1$. We expect $S_{\delta\sigma}(k)$ to capture local composition fluctuations better than $S_{\sigma}(k)$.

The structure factor $S_{\delta\sigma}(k)$ is shown in Figure~\ref{fig:skvolume}(b). Like the full diameter structure factor, $S_{\delta\sigma}(k)$ shows only mild changes as a function $\phi$. In contrast to $S_\sigma(k)$, however, $S_{\delta\sigma}(k)$ presents a more complex pattern and a marked suppression around wave numbers $k^* \approx 7$, corresponding to typical length scales of particles of intermediate sizes. Superficially, this dip might indicate an anticorrelation between diameter fluctuations over lengths of order $2\pi/k^*$. This, in turn, suggests the presence of local compositional order involving particles of different sizes, similar to the chemical ordering known in simple binary mixtures~\cite{bhatia1970structural}. This dip gets more pronounced  as $\phi$ increases, but its depth varies smoothly, see Figure~\ref{fig:skvolume}(d). In addition, the smooth evolution of these structure factors does not seem to correlate with the evolution of the glassy behavior of the system. Note finally that the fluctuations observed at the smallest wave number compatible with the simulation cell are not systematic and are due to statistical noise.

In polydisperse hard spheres, fluctuations of local volume fraction do not occur only via variation of the local number density, but can also be mediated by size dispersity. An appropriate correlation function to capture these fluctuations is the local volume structure factor $S_v(k)$, obtained by setting $w_i=v_i=4 \pi R_i^3/3$, where $R_i=\sigma_i/2$.
We computed $S_v(k)$ and found that its overall shape is similar to that of $S_\sigma(k)$, except at small $k$ (not shown). 
In this regime, $S_v(k)$ behaves asymptotically as the spectral density $\chi(k)$~\cite{wu2015search}, for which $w_i = \frac{4 \pi}{k^3} \left( \sin(k R_i) - (kR_i) \cos(kR_i) \right)$ in three dimensions. This latter quantity, which provides direct insight into hyperuniform behavior in jammed packings~\cite{zachary2011hyperuniformity,wu2015search}, also evolves smoothly by increasing density in equilibrium polydisperse hard spheres~\cite{ozawa2017exploring}.
Thus, we conclude that fluctuations of both the local diameter and of the local volume fraction evolve gradually with volume fraction and they reveal only weak compositional order. This conclusion is consistent with the very smooth evolution that we have observed of partial structure factors and pair correlation functions obtained by discretizing the particle size distribution into discrete families (not shown). The overall conclusion of this analysis of compositional order is that the present system is a good glass-former that presents very weak fluctuations of the composition, and remains amorphous and well-mixed even in deep supercooled states. 

\begin{figure*}[t]
\centering
\includegraphics[width=.95\linewidth]{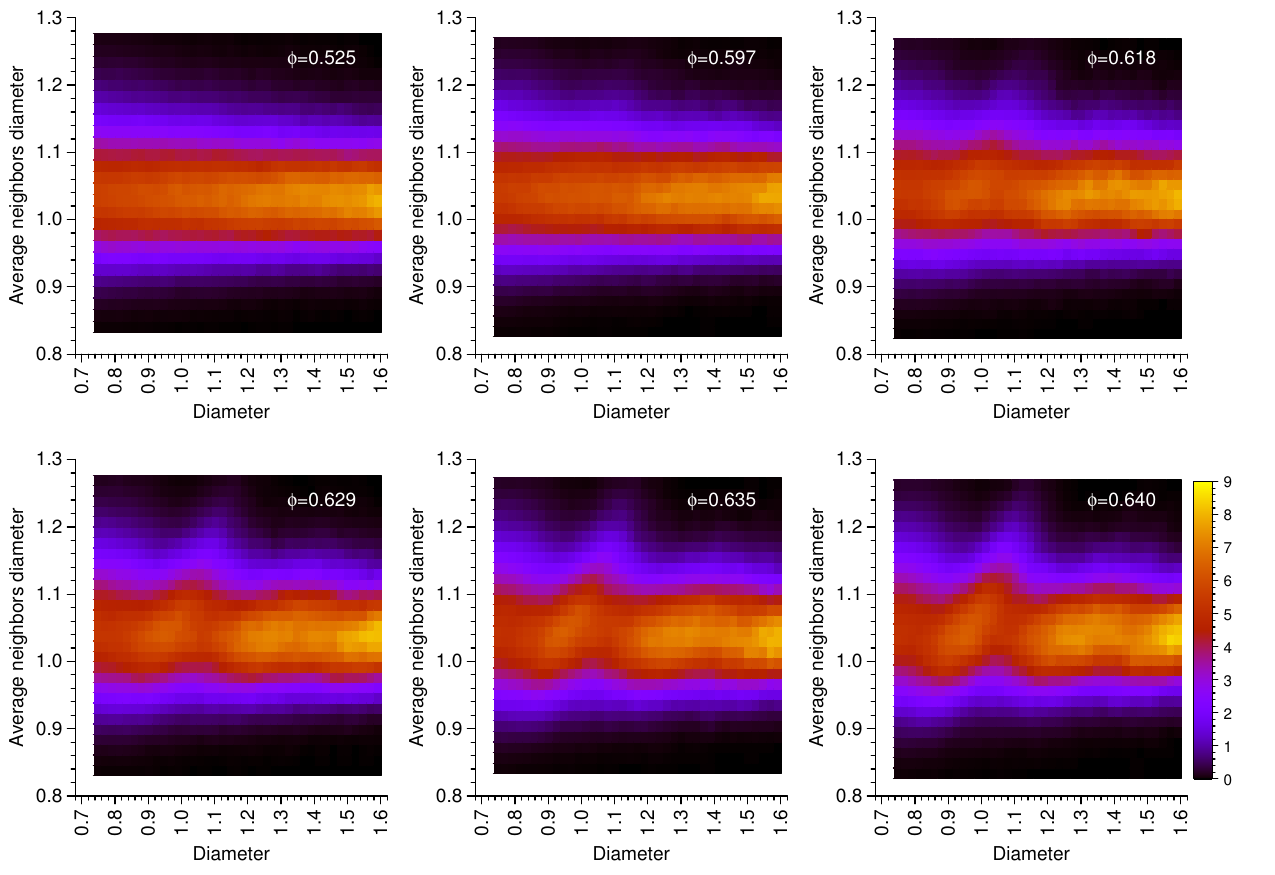}
\caption{Conditional probability distribution $P(\tilde{\sigma}|\sigma)=P(\tilde{\sigma}, \sigma)/P(\sigma)$ for several volume fractions, as indicated by the labels. 
}
\label{fig:comp}
\end{figure*}

We now look for an order parameter that detects more precisely fluctuations of local compositional order in the neighborhood of a given particle. Specifically, we consider fluctuations of the diameter within the first shell of neighbors. The neighbors of a particle are obtained from the radical Voronoi tessellation, see Section~\ref{sec:methods}. We introduce the average neighbor diameter $\tilde{\sigma}(i)$ of the $i$-th particle
\begin{equation}
 \tilde{\sigma}(i) = \frac{1}{n_{\rm b}(i)} \sum_{j=1}^{n_{\rm b}(i)} \sigma_j,
\label{eq:def_R_bar} 
\end{equation}
where the sum runs over the $n_{\rm b}(i)$ neighbors of the $i$-th particle. We then compute the joint probability distribution for $\tilde{\sigma}$ and $\sigma$, $P(\tilde{\sigma} , \sigma)$. To account for the polydispersity of the system, we actually focus on the conditional probability distribution $P(\tilde{\sigma}|\sigma)=P(\tilde{\sigma} , \sigma)/P(\sigma)$. This distribution is shown in Figure~\ref{fig:comp} for various volume fractions. At small and intermediate volume fractions, the average neighbor diameter is essentially independent of $\sigma$.
This confirms that, at least for not too dense conditions, smaller particles tend to be surrounded on average by larger ones and vice versa. This chemical local ordering is however only apparent, as it simply means that each particle (small and big) feels the same mean-field environment.

For volume fraction above $\phi\approx 0.6$, however, the distribution $P(\tilde{\sigma} | \sigma)$ presents an additional feature at intermediate values of $\sigma$. In the range $0.8<\sigma<1.1$, we clearly see a spot displaying an excess of positive correlation between $\sigma$ and $\tilde{\sigma}$, which becomes more marked with increasing $\phi$. We found that a similar excess correlation is also visible when the central particle is included in the definition of $\tilde{\sigma}(i)$ in Eq.~(\ref{eq:def_R_bar}) (not shown). A possible interpretation of this excess correlation is that the system presents local arrangements that involve particles of similar sizes, forming more regular and symmetric structures. We will show in the next section that this feature is due to the appearance of icosahedral structures, which provide the most regular local arrangements at high density.

\subsection{Geometric order}
\label{sec:geo}

Recent numerical and experimental studies provide evidence of preferred geometric motifs in simple glass-formers~\cite{leocmach2012roles,coslovich2007understanding,royall2017locally}. These motifs include local icosahedral structures, which are the structural building block of some metallic glass-formers~\cite{hirata2013geometric}, but also polytetrahedral structures~\cite{Anikeenko_Medvedev_2007} or compositionally frustrated local crystalline structures~\cite{DellaValle_Gazzillo_Frattini_Pastore_1994,Crowther_Turci_Royall_2015}. One important and delicate question is to what extent these locally favored motifs correlate over larger length scales. Malins {\it et al.}~\cite{malins_identification_2013,malins2013lifetimes} have analyzed the structure factors associated to locally favored structures in two Lennard-Jones mixtures. The results of~\cite{malins_identification_2013} indicate that icosahedral domains are weakly correlated, even though the concentration of icosahedral structures is high enough that the domains percolate. Recent works~\cite{charbonneau2013decorrelation,Wyart_Cates_2017} have further emphasized that static correlations are essentially local in the temperature regime accessible to conventional simulations.
In this section, we analyze the preferred local order of the polydisperse model and quantify its spatial extent over a very broad range of supercooling.

\begin{figure}[!t]
  \centering
  \includegraphics[width=\linewidth]{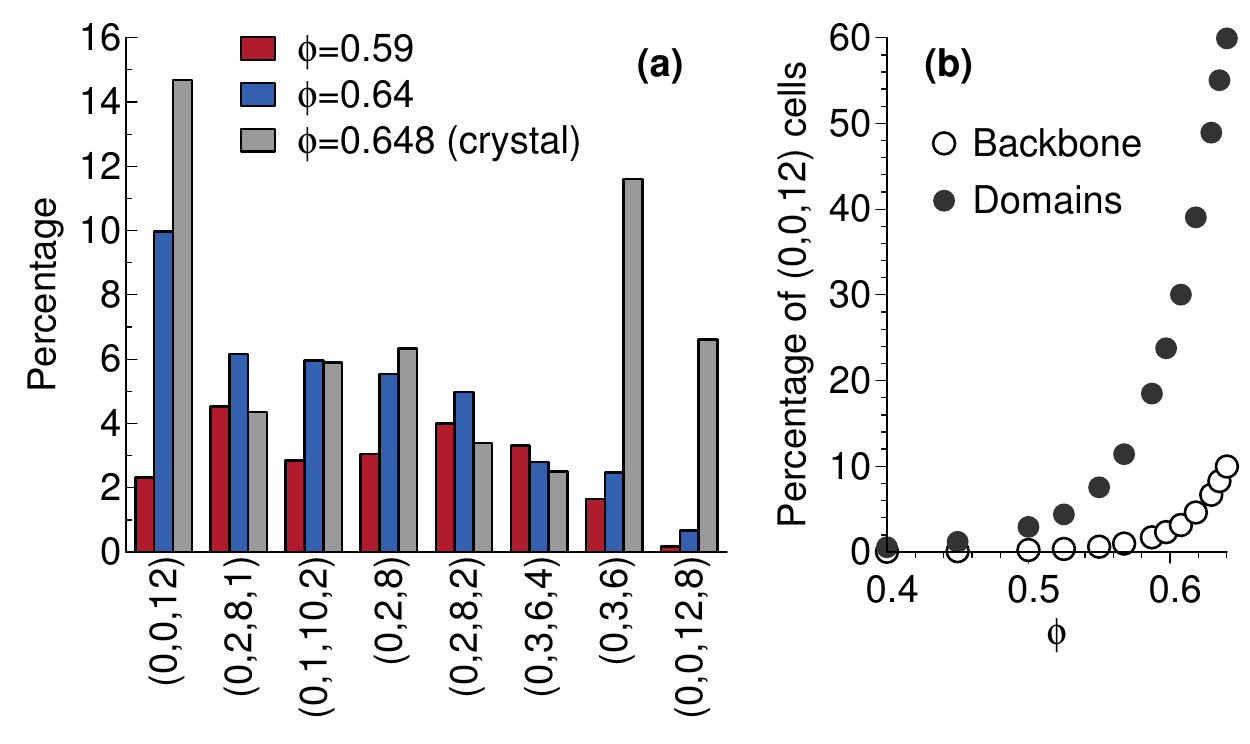}
  \caption{(a) Percentage of most frequent Voronoi cells for two selected volume fractions, and for the sample which partially crystallized. (b) Fraction of particles forming the backbone of icosahedra, i.e. at the center of a (0,0,12) cell, (empty points) and belonging to icosahedra domains, i.e. at the center or on the vertices of a (0,0,12) cell, (filled points) as a function of volume fraction.}
\label{fig:lfs}
\end{figure}

In Figure~\ref{fig:lfs} we show the fraction of most frequent Voronoi cells found around the mode-coupling crossover ($\phi_\textrm{mct}\approx 0.6$) and at the largest volume fraction $\phi=0.64$. We also include results for a smaller sample ($N=1000$) that partially crystallized during our swap MC simulations at a slightly larger volume fraction ($\phi=0.648$). We will discuss in detail the structural features of the partially crystallized sample further below, see Section~\ref{sec:xtal}. We see that icoshaedral structures, associated to $(0,0,12)$ Voronoi cells, are the most frequent ones beyond the crossover volume fraction $\phi_{\rm mct}$ and amount to about $10\%$ of the total number of cells at large volume fractions. We find that at the largest $\phi$ about $60\%$ of particles are involved in icosahedral domains, i.e. either being at the center or at the vertices of a $(0,0,12)$ cell. The fraction of particles at the center of a (0,0,12) cell and that of the particles involved in icosahedral domains is shown in Figure~\ref{fig:lfs}(b) as a function of $\phi$. Both quantities increase steadily with increasing $\phi$.

To quantify the spatial correlations associated to icosahedral order, we introduce a microscopic field $w_i$ which equals $1$ if the $i$-th particle belongs to an icosahedral structure and $0$ otherwise. We further distinguish between icosahedral backbones and icosahedral domains. For the former, $w_i=1$ only if the $i$-th particle is at the center of a $(0,0,12)$ cell. For the latter, $w_i=1$ if the $i$-th particle is the center of a $(0,0,12)$ cell or at the vertices of a $(0,0,12)$ cell. In contrast to~\cite{malins_identification_2013}, we normalize the corresponding structure factors $S_{\rm b}(k)$ and $S_{\rm d}(k)$ by the average number of particles forming icosahedral backbones and domains, respectively, and not by $N$. This is done to remove the trivial part of the state dependence of the correlation functions.

\begin{figure}[tb]
  \centering
  \includegraphics[width=\linewidth]{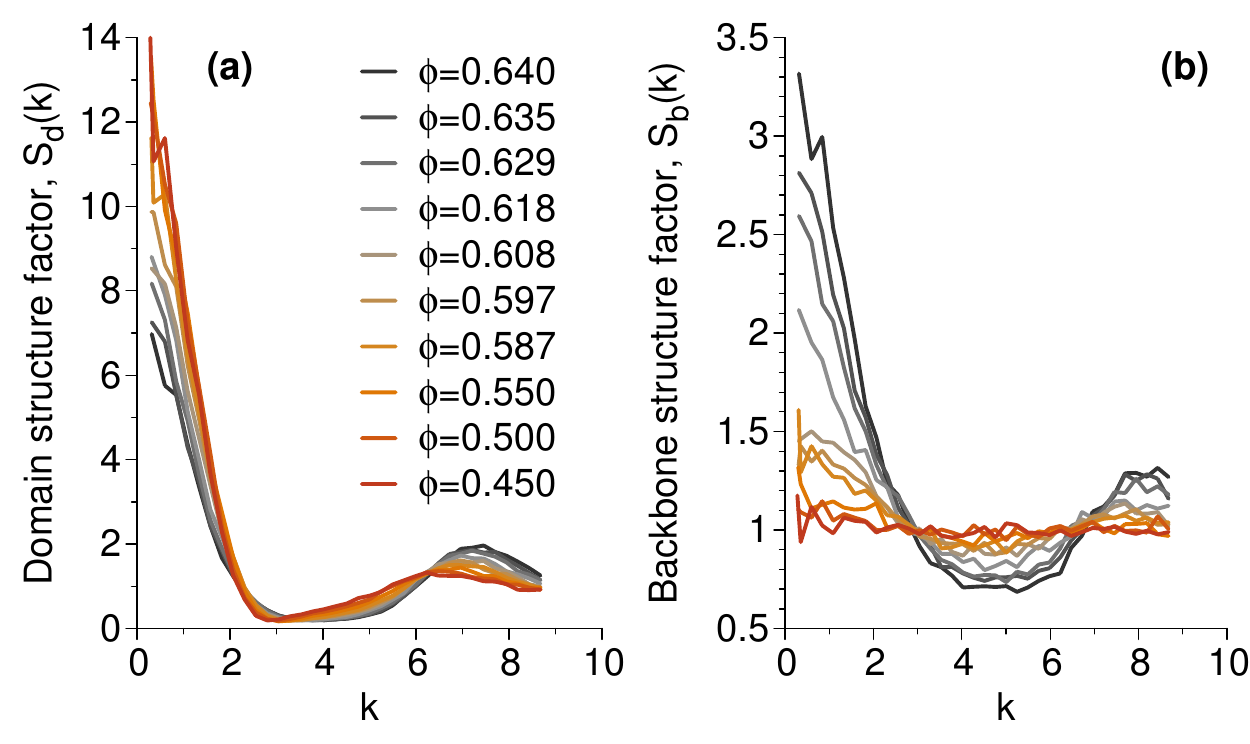}
  \caption{Structure factors of (a) the icosahedral domains $S_{\rm d}(k)$ and (b) the icosahedral backbone $S_{\rm b}(k)$ for several volume fractions.}
\label{fig:skico}
\end{figure}

\begin{figure}[tb]
  \centering
  \includegraphics[width=\linewidth]{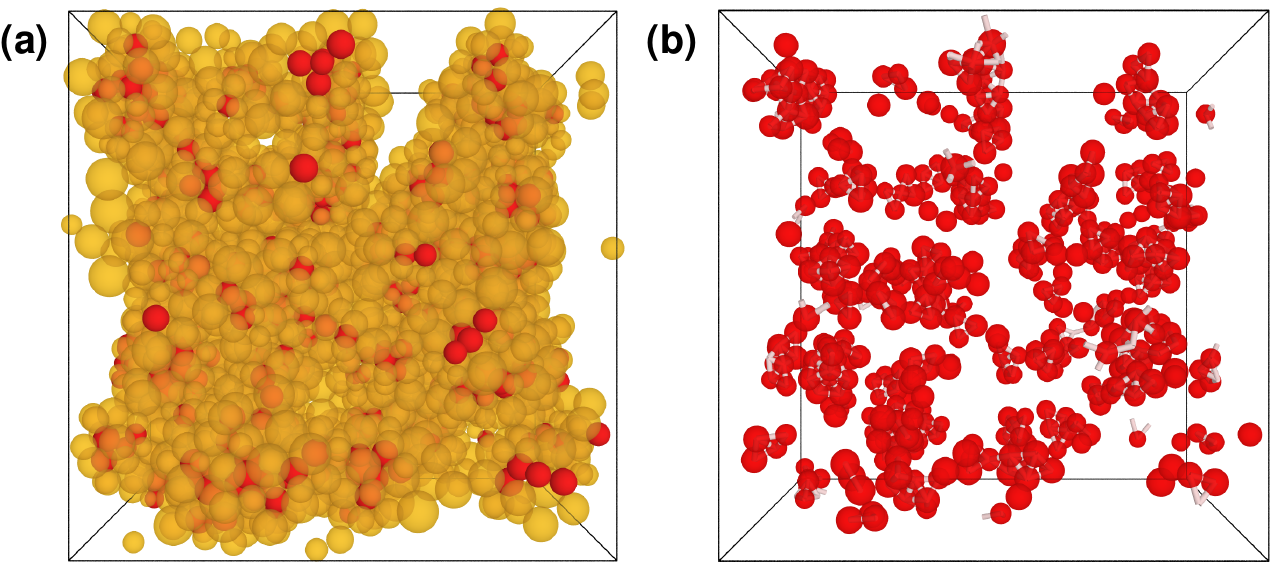}
  \caption{(a) Particles in icosahedral domains for a configuration at $\phi=0.64$. Red particles are at the center of a (0,0,12) cell, orange ones are at the vertices. (b) Icosahedral backbones of the configuration in (a). The white bonds connect neighbors centers of (0,0,12) cells.}
\label{fig:icosnap}
\end{figure}

The resulting structure factors are shown in Figure~\ref{fig:skico}. The domain structure factor $S_{\rm d}(k)$ shows a peak at $k=0$ at any density (including the unstructured, non-glassy fluid at moderate density), whose height decreases with increasing $\phi$. This indicates that the icosahedral domains are only weakly correlated. The presence of a peak around $k=0$ thus merely reflects the icosahedral form factor but not a nontrivial large-scale correlation. By contrast, correlations in the icosahedral backbone are nearly absent at low density and start to increases markedly beyond $\phi \approx 0.60$. The icosahedral backbone thus reveals subtle but nontrivial changes in the structure of the fluid. Typical snapshots of icosahedral domains and backbones at high density are shown in Figure~\ref{fig:icosnap}(a) and (b), respectively.

\begin{figure}[t]
  \centering
  \includegraphics[width=\linewidth]{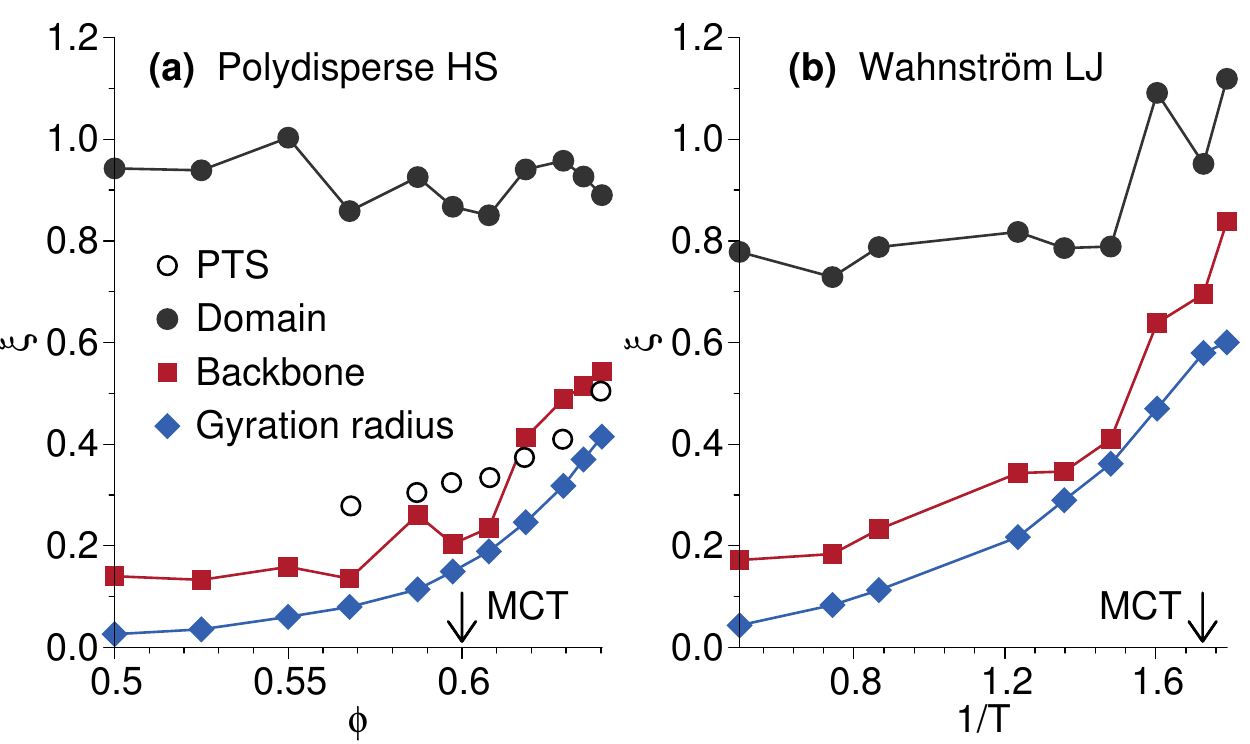}
  \caption{Length scales of icosahedral order. Domain correlation length $\xi_{\rm d}$, backbone correlation length $\xi_{\rm b}$ and backbone gyration radius $R_{\rm g}$ as a function of (a) $\phi$ in the polydisperse system and (b) $1/T$ in the Wahnstr\"om  mixture. Open circles in panel (a) represent the scaled PTS length $\xi_{PTS}/7$ as obtained in Ref.~\cite{berthier2017configurational}. The vertical arrows indicate the location of the MCT crossover.}
\label{fig:lengths}
\end{figure}

To determine the correlation length associated to icosahedral domains and backbones, we fitted to the low $k$ portion of the structure factors $S_{\rm d}(k)$ and of $S_{\rm b}(k)$, respectively, using the Ornstein-Zernicke function,
\begin{equation}
S_{\alpha}(k) = \frac{S_{\alpha}(0)}{1+(\xi_{\alpha} k)^2},
\end{equation}
where $\alpha=$ d or b.
We restricted our fits to $k<2.2$. We checked that the trends found using this approach are consistent with those obtained by manually rescaling the structure factors so as to optimize data collapse at small $k$. In Figure~\ref{fig:lengths} we show the variation of the correlation lengths $\xi_{\alpha}$ as a function of $\phi$ for both icosahedral domains and backbones. The domain correlation length remains approximately constant around $1$ interparticle distance throughout the studied range of volume fraction. This confirms that the peak observed around $k=0$ in $S_\textrm{d}(k)$ has a trivial origin. By construction, the backbone correlation length $\xi_\textrm{b}$ is smaller than $\xi_\textrm{d}$, but it increases markedly (by about a factor 3) upon increasing $\phi$ beyond the mode-coupling crossover density. The maximal value reached, $\xi_\textrm{d} \sim 0.5$, remains however quite modest and this small growth of the correlation length should be contrasted with the striking visual impression provided by the snapshot in Figure~\ref{fig:icosnap}(a) where the configuration appears full of icosahedral structures.
For comparison, we also include the static point-to-set~\cite{Bouchaud_Biroli_2004,Biroli_Bouchaud_Cavagna_Grigera_Verrocchio_2008} correlation length obtained in Ref.~\cite{berthier2017configurational}, scaled to roughly match the backbone length around the MCT crossover. We see that the relative increase of $\xi_{PTS}$ over the studied range of densities is qualitatively similar to the one of the icosahedral backbone. It would be interesting to further investigate the connection between order agnostic correlations, such as point-to-set correlations, and locally favored structures, as already suggested in Refs.~\cite{hocky_correlation_2014,Russo_Tanaka_2015,Yaida_Berthier_Charbonneau_Tarjus_2015}.

A different approach to characterize the extent of icosahedral order is to compute the gyration radius
\begin{equation}
R_{\rm g} = \left( \frac{1}{M} \sum_{j=1}^{M} (\textbf{r}_j-\textbf{r}_\textrm{b})^2\right)^{1/2}, 
\end{equation}
where $\textbf{r}_\textrm{b} = \frac{1}{M} \sum_{j=1}^{M} \textbf{r}_j$.
$M$ is the number of connected icosahedral particles, see~\cite{malins_identification_2013}. The results for the backbone gyration radius are included in Figure~\ref{fig:lengths}. $R_{\rm g}$ increases by increasing $\phi$ following the trend of the correlation function. We also found that the domain gyration radius tends to substantially overestimate the correlation length, consistent with the results of~\cite{malins_identification_2013}. We note that $R_{\rm g}$ is not well defined as soon as the cluster percolates through the system and therefore we do not show these results here. Moreover, percolation of icosahedral domains has no obvious connection with the (swap) dynamics of the system, which evolves smoohtly throughout the studied temperature regime~\cite{berthier2017configurational}.

It is interesting to compare the behavior of the model at hand with one of the Wahnstr\"om Lennard-Jones mixture~\cite{wahnstrom}, which is a binary glass-former displaying a fairly strong icosahedral ordering~\cite{coslovich2007understanding}. We computed the backbone correlation length and gyration radius for this model, using the same parameters, density and units as in~\cite{wahnstrom,coslovich2007understanding,malins_identification_2013}. Both quantities increase markedly by decreasing temperature already above the mode-coupling crossover, see Fig~\ref{fig:lengths}(b). The domain correlation length also increases slightly at sufficiently low temperature, but the absolute values of all these lengths remain small, because geometric frustration is strong in this system~\cite{Turci_Tarjus_Royall_2017}. The behavior of this Lennard-Jones mixture is thus qualitatively similar to the one of the polydisperse system, but the latter differs for two main reasons. First, the structure of the polydisperse system remains highly disordered in the moderately supercooled regime and only starts to develop some geometric order well beyond the crossover volume fraction $\phi_{\rm mct}$. Thus, the increase of the local order is shifted to a considerably deeper degree of supercooling compared to the Wahnstr\"om Lennard-Jones mixture.
Second, we recently showed~\cite{ozawa2017exploring} that icosahedral structures are actually more distorted than in simple mixtures due to local compositional fluctuations. 

\begin{figure}[!t]
  \centering
  \includegraphics[width=\linewidth]{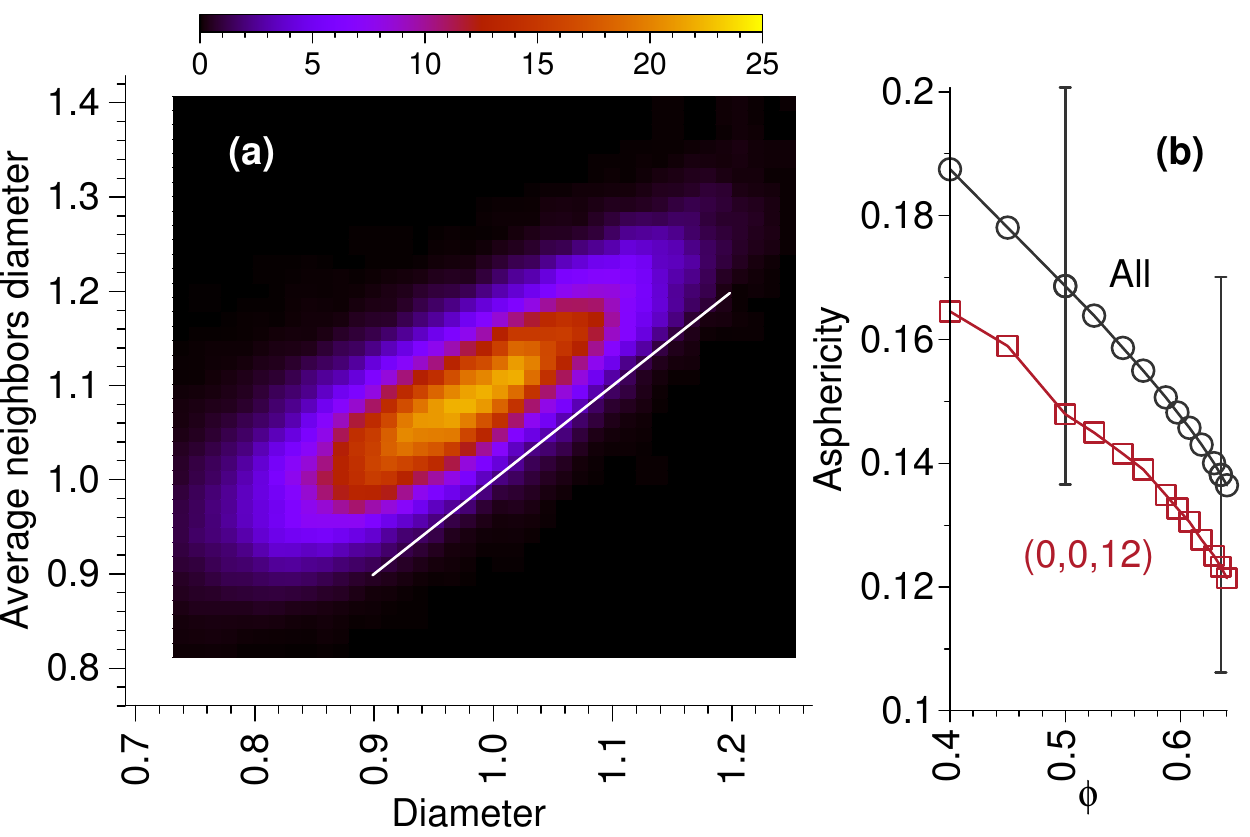}  
  \caption{
(a) Conditional probability distribution $P_\textrm{ico}(\tilde{\sigma} | \sigma)$ for icosahedra centers at $\phi=0.64$. The straight line is a guide for the eyes.
(b) Asphericity of all Voronoi cells (circles) and $(0,0,12)$ cells (squares) as a function of $\phi$.}
\label{fig:sphericity}
\end{figure}

We now shed some light on the interplay between geometric and compositional order. In Figure~\ref{fig:comp} we noticed the emergence of an excess correlation in the $P(\tilde{\sigma} | \sigma)$ distribution, which becomes increasingly visible at larger densities. We argue that this correlation is due to the growing icosahedral order. In Figure~\ref{fig:sphericity} we show the distribution $P_\textrm{ico}(\tilde{\sigma} | \sigma)$ restricted to particles at the center of an icosahedron, evaluated at the largest volume fraction. The clear correlation between $\sigma$ and $\tilde{\sigma}$ corresponds nicely to the feature observed in the full distribution $P(\tilde{\sigma} | \sigma)$. A plausible interpretation of the excess correlation in $P(\tilde{\sigma} | \sigma)$ is that icosahedra tend to be more regular and spherical than other structures. At sufficiently high density, a high degree of sphericity also likely implies that particles involved in the local structure have similar sizes, thus a correlation between $\sigma$ and $\tilde{\sigma}$. Note that this trend by itself does not imply fractionation, which should lead to a much sharper structural change.

To confirm that icosahedra are indeed the most regular structures in the model, we measure the asphericity of the Voronoi cell by computing the normalized standard deviation of the distances from the $i$-th particle 
\begin{equation}
s_i = \frac{1}{\tilde{r}_i}\sqrt{\frac{1}{n_{\rm b}(i)}\sum_{j=1}^{n_{\rm b}(i)}(r_{ij} - \tilde{r}_i)^2} ,
\end{equation}
where $\tilde{r}_i = (\sum_{j=1}^{n_{\rm b}(i)} r_{ij}) /n_{\rm b}(i)$ and $r_{ij}$ are respectively the average nearest neighbors distance from the $i$-th particle and the distance between particles $i$ and $j$. We expect this measure to be closely related to other measures of regularity (i.e., tetrahedricity) previously introduced in the study of simple particulate systems~\cite{krekelberg2006model,anikeenko2008structural}. First, we observe from Figure~\ref{fig:sphericity}(b) that the average asphericity of the Voronoi cells decreases as the system gets denser, as expected. We find that icosahedra are appreciably more regular than the other structures. By restricting ourselves to the most spherical structures, we find that the proportion of icosahedra is significantly higher than in the bulk. Specifically, we computed the Voronoi cell statistics for the $2\%$ most spherical particles at $\phi=0.64$. We find that $28\%$ of these highly spherical structures are icosahedral, which should be contrasted to their bulk average of about $10\%$. Among the highly spherical structures, the proportions of all other main signatures are lower than in the bulk. This confirms that icosahedral structures, despite the enhanced compositional disorder~\cite{ozawa2017exploring}, still provide the most regular and spherical arrangements in the system.

\subsection{Partial crystallization}
\label{sec:xtal}

The structural analysis carried out so far concerns equilibrium, disordered fluid states. However, like any supercooled fluid, the model at hand is thermodynamically metastable and sufficiently long simulations with the swap Monte Carlo algorithm may trigger a fluctuation towards the ground state. In systems with sufficiently high polydispersity, reaching the crystalline ground state may involve fractionation into families of particles characterized by similar diameters~\cite{sollich2010crystalline}. However, this process may take extremely long times, and alternate crystallization scenarios, not involving fractionation, have also been observed~\cite{fernandez2007phase,fernandez2010separation}.

\begin{figure}[t]
  \centering
  \includegraphics[width=\linewidth]{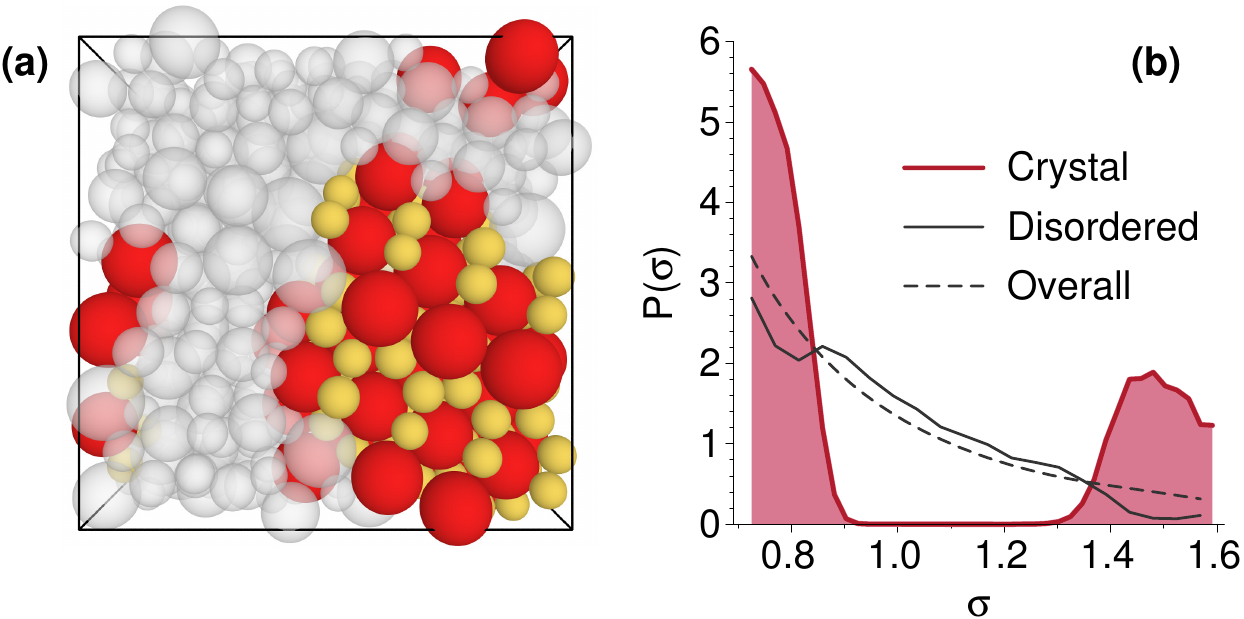} 
  \caption{(a) Snapshot of a partially crystallized sample at $\phi=0.648$. Red and yellow particles are centers of (0,0,12,8) and (0,3,6) cells, respectively, and define the crystalline region of the sample. The remaining particles constitute the disordered portion of the sample and are shown as transparent white spheres. (b) Distribution of the diameter $P(\sigma)$ in the crystalline region (red curve with shaded area), in the disordered region (black curve), and in the bulk (dashed curve).
\label{fig:xtal}}
\end{figure}

In this section, we focus on a sample of $N=1000$ particles at a volume fraction $\phi=0.648$, which partially crystallized during the course of a long enough Monte Carlo swap simulation. Crystallization was easily detected by an anomalous behavior of the pressure and dynamic behavior, as well as by visual inspection of the particles configuration. We have found a few of such crystallization events during the course of our studies, but these only happened at very large densities ($\phi > 0.645$), i.e., beyond the range of volume fractions corresponding to the laboratory glass transition ($\phi=0.635-0.645$~\cite{berthier2017configurational}), and after long simulations times, see below. These samples were excluded of all published analysis of dense supercooled liquid states~\cite{berthier2016equilibrium,berthier2017configurational}.
In the following, we provide some quantitative insight into the structure of the partially crystallized sample and highlight the differences compared to the normal fluid states.
Note that from the viewpoint of glass transition studies, these crystallization events are only problematic when they occur on time scales comparable to the structural relaxation time $\tau_\alpha$, which controls the equilibration of density fluctuations. In a related study we have shown that it is possible to alleviate this problem by introducing non-additive pair interactions, which help to suppress crystallization events even further~\cite{ninarello2017}.

Visual inspection of the particles configurations in Figure \ref{fig:xtal}(a) immediately shows that the system phase separates into a disordered and a crystalline region. We found that particles in the crystalline region are clearly associated to $(0,3,6)$ and $(0,0,12,8)$ Voronoi cells. The proportions of these cells can thus be used as a marker of the system instability, since they increase markedly upon partial crystallization, see Figure~\ref{fig:lfs}. Particles located at the center of $(0,3,6)$ and $(0,0,12,8)$ cells are the smallest and the largest particles, respectively, and are highlighted accordingly in Figure~\ref{fig:xtal}(a). Thus, while the disordered region has a local polydispersity similar to the one of the homogeneous fluid, the crystalline one comprises only a subset of the particles, and is completely devoid of particles of intermediate sizes. This is demonstrated in Figure~\ref{fig:xtal}(b), where we compare the overall diameter distribution $P(\sigma)$ to the one measured in the crystalline region, i.e. for particles at the center of either $(0,3,6)$ or $(0,0,12,8)$, and in the disordered region.

\begin{figure}[t]
  \centering
  \includegraphics[clip,trim=0 25 0 0,width=0.9\linewidth]{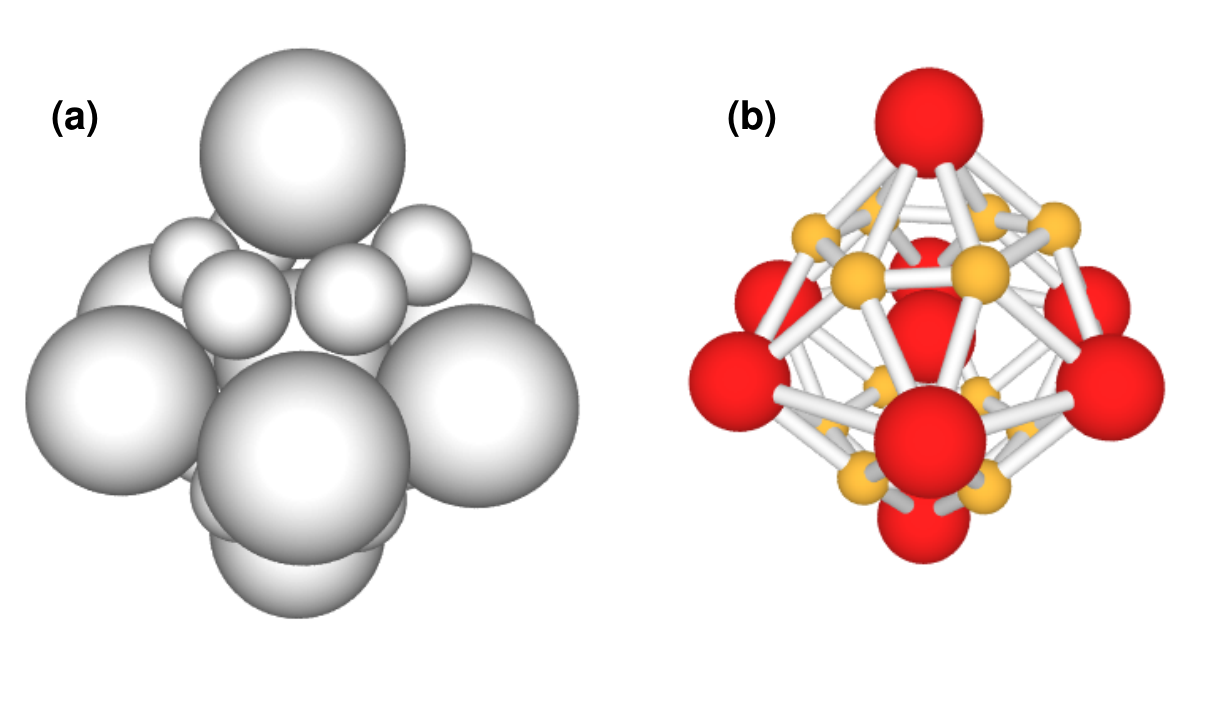} 
  \caption{Structure of a representative (0,0,12,8) cell found in the crystalline region. In panel (a) spheres are drawn to scale. In panel (b) spheres are scaled to half of their size and bonds are added between neighboring particles to highlight the hexagonal symmetry of the crystal. In (b) small and large particles are shown as yellow and red spheres, respectively.\label{fig:snap}}
\end{figure}

The symmetry of the crystal is that of aluminum diboride, AlB$_2$, with small and large particles playing the role of B and Al atoms, respectively. The crystal structure has an hexagonal symmetry and is formed by interleaved layers of small and large particles. The typical shape of the first coordination shell around a large particle of the crystal is illustrated in Figure~\ref{fig:snap}, where we show the structure of a (0,0,12,8) Voronoi cell. The typical size ratio $\gamma\approx 0.5$ between the small and large particles forming the crystal is close to the one ($\gamma = 0.58$) of AlB$_2$-forming binary hard colloids~\cite{Bartlett_Ottewill_Pusey_1992} and lies in the stability range ($0.42\le \gamma \le 0.59$) expected from theoretical studies of binary hard spheres~\cite{Cottin_Monson_1995,Filion_Dijkstra_2009}. Finally, we note that the Voronoi tessellation of the AlB$_2$ lattice comprises indeed only $(0,0,12,8)$ and $(0,3,6)$ cells, centered around Al and B atoms respectively, see e.g.~\cite{Travesset_2017}.

\begin{figure}[t]
  \centering
  \includegraphics[width=\linewidth]{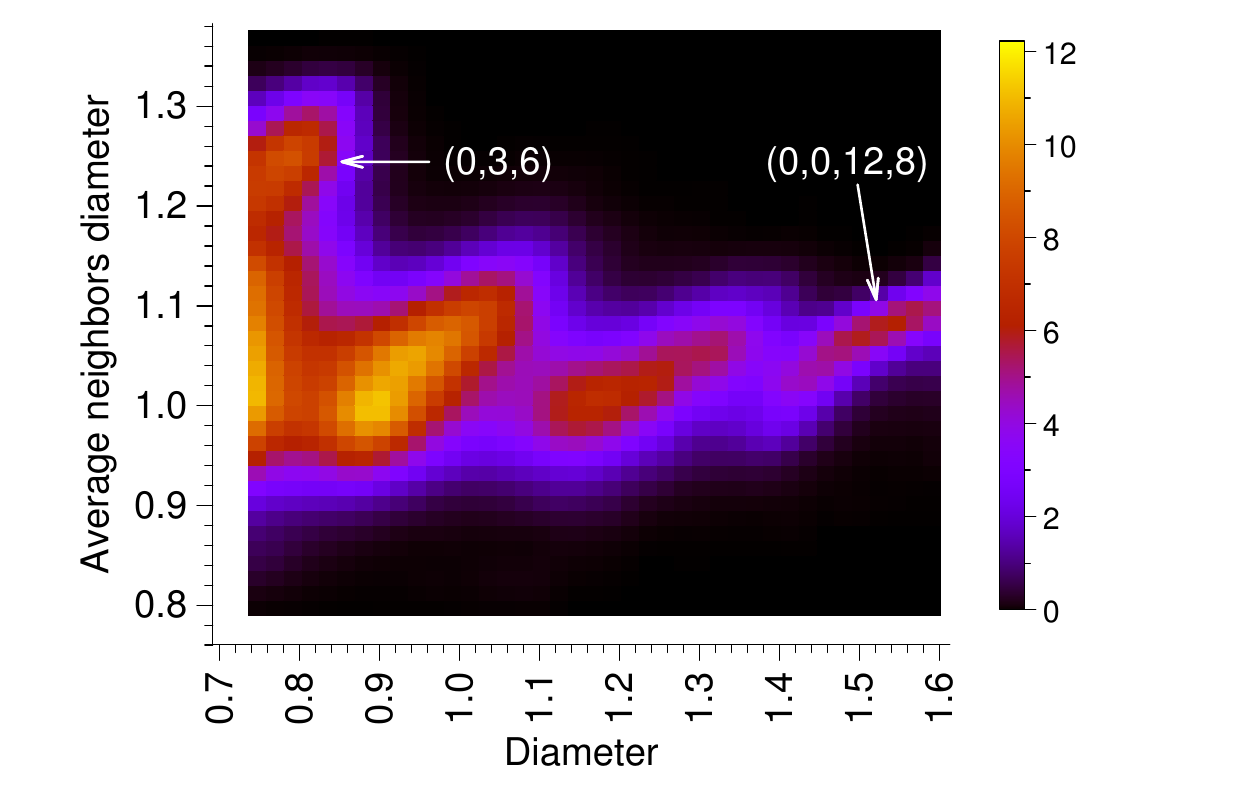}
  \caption{\label{fig:xtal2} Joint probability density $P(\tilde{\sigma},\sigma)$ for the partially crystallized sample at $\phi=0.648$. The spots indicated by the arrows are associated to the crystallized portion of the sample, and are characterized by Voronoi cells with the indicated signature.}
\end{figure}

To investigate the structure of the partially crystallized sample more quantitatively, we compute the distribution of the compositional order parameter $(\tilde{\sigma}, \sigma)$. In contrast to the equilibrium fluid states studied in the previous sections, the partially crystallized sample displays a complex distribution $P(\tilde{\sigma}, \sigma)$ characterized by multiple spots. Two of these spots, marked by arrows in Figure~\ref{fig:xtal2}, are clearly associated with the crystalline region of the sample and involve the smallest and the biggest particles in the sample. Note that, to enhance visualization, we show here the full distribution $P(\tilde{\sigma}, \sigma)$ instead of $P(\tilde{\sigma}|\sigma)$. By computing the Voronoi statistics restricted to specific ranges of $\sigma$ and $\tilde{\sigma}$, we confirmed that the crystalline spots indicated by arrows in Figure~\ref{fig:xtal2} correspond to $(0,3,6)$ and $(0,0,12,8)$ Voronoi cells. Note, however, that some of the smallest particles have values of $\tilde{\sigma}$ comparable to the ones found in fluid states. Visual inspection of the particle configurations indicates that these small particles populate the disordered portion of the sample.

The distribution $P(\tilde{\sigma},\sigma)$ displays two additional spots located at intermediate values of $\sigma$ and characterized by a distinct positive correlation. One of these spots, between $0.8<\sigma<1.1$, is associated to icosahedral structures and is similar to the one found in the fluid samples. By contrast, no clear structural signature stands out in the second spot between $1.1<\sigma<1.35$~\footnote{The most frequent signatures in this spot are $(0,0,12,5)$ and $(0,1,10,6)$, with $14\%$ and $12\%$ respectively.}. Surprisingly, we found that particles that contribute to the two central spots of $P(\tilde{\sigma},\sigma)$ are not spatially segregated from one another and are characterized by very similar local polydispersities~\footnote{The local polydispersities of the two spots are in the range 20--21\%. They were measured by computing the distribution of diameters of particles in the corresponding range of $\sigma$ and of their reespective neighbors.}. Thus, the existence of a positive correlation between $\sigma$ and $\tilde{\sigma}$ does not imply \textit{per se} fractionation and may be attributed instead to a subtle local geometric ordering. Overall, our results confirm that, in practice, crystallization and phase separation in polydisperse hard spheres follow a more complex pattern than predicted by existing theoretical models~\cite{sollich2010crystalline}.

Finally, we estimated the crystallization time by performing additional simulations of independent samples of 1000 particles at several volume fractions. Crystallization was detected by inspecting the evolution of the percentage of (0,0,12,8) cells, which typically fluctuates between 0.5\% and 2\% for fluid states and exceeds 5\% in partly crystallized samples. For $\phi=0.635$ and $\phi=0.643$ no crystallization events were observed during simulations covering 600$\tau_\alpha$ and 200$\tau_\alpha$, respectively, where $\tau_\alpha$ is the structural relaxation time measured from the self intermediate scattering function~\cite{berthier2017configurational}. Thus, in this density regime and for this system size, our swap Monte Carlo simulations can safely probe the structure of the metastable equilibrium fluid. For $\phi=0.648$, two independent samples out of 10 crystallized during simulations of about $\tau_x=6\times 10^8$ Monte Carlo steps, corresponding to about $30\tau_\alpha$. A conservative upper bound to the crystallization rate $1/(\tau_x V)$ is thus $3\times 10^{-12}$ in reduced units~\footnote{The time unit is given by one Monte Carlo step, comprising $N$ attempts to either displace a particle or swap the identities of a pair $(i,j)$ of particles such that $|\sigma_i-\sigma_j|<0.2$.}. We emphasize that this value may still be strongly affected by finite size effects induced by the periodic boundary conditions.

\section{Conclusions}
\label{sec:conclusions}

We characterized the local structure of a fluid of polydisperse hard spheres over a wide range of volume fractions, where conventional computer simulations fail to equilibrate.
We showed that local compositional order increases smoothly with increasing the volume fraction, with little correlation with the glassy evolution of the system. Concomitantly, local geometric order associated to icosahedral particle arrangements grows steadily at very large volume fractions.
We extracted correlation lengths associated to icosahedral structures using weighted structure factors and the gyration radius. These correlation lengths increase appreciably only at large packing fractions and their absolute values remain small over the entire glassy regime we could probe. It is interesting to compare this behavior with the results for the Wahnstr\"om Lennard-Jones mixture, which is a glass-forming model displaying a large amount of icosahedral structures. The measured growth of icosahedral length scales is qualitatively similar in the two models, but the glassy regime explored in the Wahnstr\"om mixture is much narrower. Thus, in the regime probed by conventional simulations, the local structure of the polydisperse model appears highly disordered, as is the case in simple binary mixtures of hard~\cite{charbonneau2013decorrelation} or quasi-hard spheres~\cite{hocky_correlation_2014}. This indicates that the role of local structure is not only system-dependent~\cite{hocky_correlation_2014}, but also highly state-dependent.

Finally, we characterized the structure of a partially crystalline sample of $N=1000$ particles obtained during long simulations at a packing fraction $\phi=0.648$. We found that the system demixes into a fluid of particles with similar polydispersity as the parent system and an AlB$_2$ crystal comprising only the smallest and largest particles of the sample. We emphasize that crystallization only occurs at a packing fraction that lies beyond the estimated laboratory glass transition, in a regime where crystal growth, driven by physically realistic dynamics, would be extremely slow on observational time scales~\cite{Cavagna_2009}. The fact that swap Monte Carlo simulations involve non-physical moves, which accelerate sampling of configuration space, makes it difficult to infer the absolute glass-forming ability of the models using ordinary dynamics, but we expect that the relative trends across systems, see Ref.~\cite{ninarello2017}, will be preserved. Investigations to tackle these issues are currently under way. Comparing the glass-forming ability of these models with experimental systems represents a major challenge that is left for future investigations.

Continuously polydisperse systems can be regarded as an extreme case of multicomponent mixtures. At first glance, these systems may appear peculiar~\cite{lubchenko2017aging}, for instance because of their formally infinite mixing entropy~\cite{frenkel2014colloidal,ozawa2017does}. However, their glassy phenomenology strongly resembles the one of conventional glass-formers~\cite{berthier2017configurational}. Moreover, our detailed structural characterization demonstrates that the local structure of polydisperse system at hand shows qualitatively similar features as other representative glass-formers such as colloidal~\cite{leocmach2012roles} and metallic glasses~\cite{hirata2013geometric}, which also display growing icosahedral order.
Thus, overall, our results confirm that continuous polydisperse systems can be regarded as good models to study the glass transition. Whether more general classes of glass-formers, such as molecular or polymeric liquids, display or not a pronounced local geometric order remains an open question to be addressed in future numerical studies.

\begin{acknowledgments}
We thank G. Tarjus for enduring discussions on the role of icosahedral order and its length scales. The research leading to
these results has received funding from the European Research
Council under the European Unions Seventh Framework 
Programme (No. FP7/2007-2013)/ERC Grant Agreement No.306845. 
This work was supported by a grant from the Simons
Foundation (No. 454933, Ludovic Berthier).
Data relevant to this work have been archived and can be accessed at \href{https://doi.org/10.5281/zenodo.1183325}{{https://doi.org/10.5281/zenodo.1183325}}.
\end{acknowledgments}

\renewcommand{\doi}[1]{\href{http://dx.doi.org/#1}{doi:\discretionary{}{}{}#1}}
\bibliography{paper}

\end{document}